\documentclass[twocolumn,showkeys,superscriptaddress]{revtex4}
\usepackage{graphicx}
\usepackage{epsfig}
\usepackage{amsmath}
\usepackage{amssymb}
\usepackage[ps2pdf,bookmarks,urlcolor=red,citecolor=blue,linkcolor=blue,
               pagecolor=blue,hyperindex,colorlinks,hyperfigures,
        ]{hyperref}

\begin{document}
\title{Dynamic polarizability of rotating particles
in electrorheological fluids}

\author{J. J. Xiao\footnote{Electronic mail: jjxiao@phy.cuhk.edu.hk \\
Present address: Department of Physics, The Hong Kong University of Science
and Technology, Clear Water Bay, Kowloon, Hong Kong, China}} \affiliation
{Department of Physics, The Chinese University of Hong Kong, Shatin, New
Territories, Hong Kong, China}

\author{J. P. Huang}
\affiliation{Surface Physics
Laboratory (National Key Laboratory) and Department of Physics, Fudan
University, Shanghai 200433, China}

\author{K. W. Yu\footnote{Electronic mail: kwyu@phy.cuhk.edu.hk}}
\affiliation{Department of Physics, The Chinese University of Hong
Kong, Shatin, New Territories, Hong Kong, China}

\affiliation{Institute of Theoretical Physics, The Chinese
University of Hong Kong, Shatin, New Territories, Hong Kong, China}

\begin{abstract}
A rotating particle in electrorheological (ER) fluid leads to a displacement
of its polarization charges on the surface which relax towards the external
applied field ${\bf E}_0$, resulting in a steady-state polarization at an
angle with respect to ${\bf E}_0$. This dynamic effect has shown to affect
the ER fluids properties dramatically. In this paper, we develop a dynamic
effective medium theory (EMT) for a system containing rotating particles of
finite volume fraction. This is a generalization of established EMT to
account for the interactions between many rotating particles. While the
theory is valid for three dimensions, the results in a special two
dimensional configuration show that the system exhibits an off-diagonal
polarization response, in addition to a diagonal polarization response, which
resembles the classic Hall effect. The diagonal response monotonically
decreases with an increasing rotational speed, whereas the off-diagonal
response exhibits a maximum at a reduced rotational angular velocity
$\omega_0$ comparing to the case of isolated rotating particles. This implies
a way of measurement on the interacting relaxation time. The dependencies of
the diagonal and off-diagonal responses on various factors, such as
$\omega_0$, the volume fraction, and the dielectric contrast, are discussed.
\end{abstract}
\date{\today}
\maketitle

\section{Introduction}
In many aspects, an electrorheological (ER) fluid containing suspensions at
rest is quite different from an ER fluid containing suspensions subjected to
rotational motions, \cite{Block84, Hemp91,Lad88, Williams00, Hu04, Zukoski91,
Negita04, Ohsawa95} which has been referred to the dynamic effect in ER
fluid, \cite{WanPRe00, WanPRE01, Yu05} and is of great relevance in various
applications of ER fluid. Reports regarding this dynamic effect have been
extensive in the literature. \cite{HuangJPCB,Negita04Rotation,EOR,
EOR1,NegativeER, NEgative2,Tao05} Typically there are three cases \cite{Yu05}
for rotation of the suspending particles about their centers: (1) rotation
due to external torque, which, for instance, might be induced by the shear
flow in the ER fluid. \cite{HuangJPCB,Lad88, WanPRe00, WanPRE01, Yu05,
Negita04Rotation} (2) particle rotation due to a rotating external field,
which is a common phenomena utilized in electrorotation assay. \cite{EOR,
EOR1} (3) spontaneous rotation of particular particles within weak conducting
fluid in a dc field above a threshold . This is first discovered almost a
century ago, \cite{Quincke} now recognized as Quincke effect or negative ER
effect. \cite{NegativeER, NEgative2} In all these cases, the rotation axis is
favored in a special direction.

The rotational motion leads to a displacement of surface charge (not for
Quincke rotation) on the particles that mediates the interactions in the ER
system. For an isolated rotating particle, it has been shown that the
displaced surface charge on the particle due to rotation will relax towards
the external applied field  ${\bf E}_0$. \cite{WanPRe00} The result of
competition between the displacement and the relaxation is a steady-state
polarization which deviates from the applied field's direction, and depends
on the product of the rotational angular velocity $\boldsymbol{\omega}_0$ of
the particle and the relaxation time ${\tau}_0$. At the same time, the
magnitude of the steady-state polarization is reduced by a factor as compared
to the resting case. \cite{Hu04, WanPRe00} which is determined by the same
competition. As the interparticle force depends crucially on the magnitude
and direction of polarizations, the complications due to the rotational
motion were shown to be dramatic. \cite{Tao05,WanPRE01} A first attempt of
two rotating particles was achieved by Wan et al. \cite{WanPRe00} However, a
more serious attempt must be directed to a suspension of finite volume
fraction. It is therefore instructive to extend the existing theories to the
scenario of many rotating particles, as in realistic ER suspensions. In which
case, it is necessary to assess the local field effects.

However, for an individual rotating particle in such a system, e.g., see
Fig.~\ref{fig:fig1}, the relaxation of the induced surface charge on the
particle will be towards the sum ${\bf E}_{\text{total}}$ of the external
applied field ${\bf E}_0$ and the local field ${\bf E}_{\text{L}}$ due to all
other particles. Moreover, the presumable polarization ${\bf P}$ of the
particles is generally not parallel to the applied field ${\bf E}_0$
[Fig.~\ref{fig:fig1}(b)].  In a heuristic example, since the local field is
always proportional to the polarization ${\bf P}$ by virtual of the Lorentz
concept, the total field is neither along the polarization nor along the
external applied field in presence of any rotational motion. The result of
the competition between rotation and relaxation, however, must lead to the
prescribed steady-state polarization. As a result, the steady-state
polarization must be determined self-consistently and it leads to an
effective medium theory for such a system of many rotating particles.
Regarding the local field effects, the difficulty lies in the tensorial
nature of the problem. More precisely, as we mentioned, it is because the
polarization is not parallel to the applied field due to the competition
between the displaced surface charge on the particles via mechanical rotation
and the polarization's relaxation towards the applied field. Therefore, a
successful theory must accommodate these properties correctly.

In the next section, we shall put forth an effective medium theory that is
able to describe such properties of ER fluid consisting rotating suspensions
in a very general situation. Then in Sec.~\ref{sec:2D} we examine in detail a
special case of two-dimensional (2D) configuration and present some simple
analytical results which will be confirmed numerically. At the same time, we
report an interesting off-diagonal polarization response arising from the
dynamic effect. We conclude the paper with some discussions in the last
section.

\section{Dynamic effective medium theory}
\begin{figure}[t]
  \includegraphics*[clip,width=0.3\textwidth]{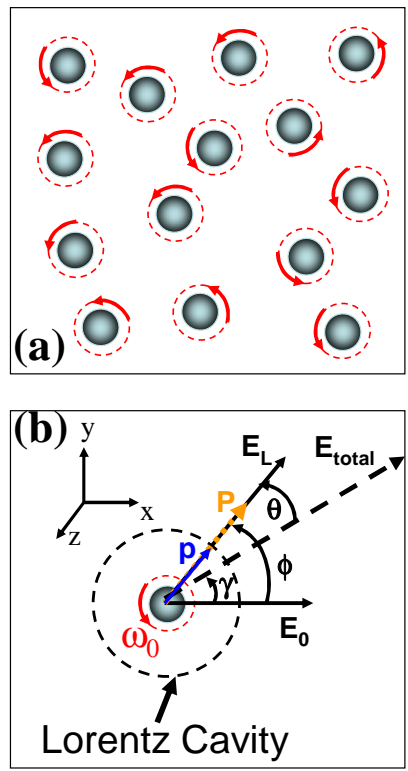}
  \caption{(Color online) (a) Cartoon representation for a
system containing rotating particles of finite volume fraction. (b)
Schematic relationship of the single particle polarization and the
self-consistent macroscopic polarization.}
  \label{fig:fig1}
  \end{figure}

Let us consider a volume fraction $\rho$ of homogeneous spherical particles
of dielectric constant $\epsilon_1$, embedded in a host medium of unit
dielectric constant [see Fig.~\ref{fig:fig1}(a)]. An external electric field
${\bf E}_0$ is applied to the system. We then assume a uniform rotational
angular velocity ${\boldsymbol \omega}_0$ for all the suspending particles.
This could be a typical ER fluid under shear, \cite{Negita04} wherein the
suspended particles are of micrometeric size and have a large ratio of
surface-to-bulk polarization.

If we first ignore the interaction between the particles, the dynamic
equation for the induced dipole moment ${\bf p}$ of a single particle is
given by \cite{WanPRe00}
\begin{equation}
\dot{\bf p} + {\boldsymbol \omega}_0 \times {\bf p} = {1\over \tau_0} \left(
{\bf p} - {\bf p}^{(0)}\right),
\end{equation}
where ${\bf p}^{(0)}$ is the dipole moment of the particle in the absence of
rotation, and $\tau_0$ is the relaxation time of the noninteracting
particles. The steady-state dipole moment of the particle can be expressed in
terms of ${\bf p}^{(0)}$
\begin{equation}
\label{eq:singlesolution} {\bf p}={{\bf p}^{(0)}+\tau_0^2
\boldsymbol{\omega}_0 (\boldsymbol{\omega}_0 \cdot {\bf
p}^{(0)})+\tau_0({\boldsymbol \omega}_0 \times {\bf p}^{(0)})\over 1 +
(\omega_0 \tau_0)^2},
\end{equation}
where $\omega_0=|\boldsymbol{\omega}_0|$. The steady-state dipole moment
${\bf p}$ of Eq.~\eqref{eq:singlesolution} is generally expressed in a sum of
axial and solenoidal terms. Furthermore, we can show that

\begin{equation}
\boldsymbol{\omega}_0 \cdot {\bf p}^{(0)}=\boldsymbol{\omega}_0 \cdot {\bf
p}.
\end{equation}
In the presence of interaction, one can define the effective dielectric
constant $\boldsymbol{\epsilon}$ in terms of the polarization ${\bf P} =
\sum_i {\bf p}_i/V$ of the system \cite{Jackson1975}
\begin{equation}
{\bf P}={\boldsymbol{\epsilon} - \boldsymbol{I} \over 4\pi} \cdot {\bf E}_0,
\label{eq:P}
\end{equation}
where $\boldsymbol{I}$ is the identity tensor. Note that by definition,
$\boldsymbol{\epsilon}$ is a tensor and frequency dependent. By invoking the
Lorentz local electric field for the steady-state responses, the induced
dipole moment is given by
\begin{equation}
{\bf p} = {\bf P} v = \boldsymbol{\alpha} \cdot \left({\bf E}_0 + {4\pi \over
3}{\bf P}\right), \label{eq:Lorentz}
\end{equation}
where $\boldsymbol \alpha$ is polarizability tensor of the rotating particle
which is also frequency dependent, and $v$ is the average volume per particle
in the suspension. By substituting Eq.~\eqref{eq:P} into
Eq.~\eqref{eq:Lorentz}, we arrive at the dynamic Maxwell-Garnett formula
\begin{equation}
\boldsymbol{\epsilon} - \boldsymbol{I} = {4\pi \over 3v} \boldsymbol{\alpha}
\cdot (\boldsymbol{\epsilon} + 2\boldsymbol{I}). \label{eq:DMGA}
\end{equation}
Thus Eq.~\eqref{eq:DMGA} represents a generalization of the static
Maxwell-Garnett approximation (MGA) to that of rotating particles. In the
absence of rotation, the polarizability tensor
$\boldsymbol{\alpha}^{(0)}=\alpha^{(0)}\boldsymbol{I}$ becomes diagonal
\begin{equation}
\alpha^{(0)} = {\epsilon_1 - 1\over \epsilon_1 + 2} a^3,
\end{equation}
where $a$ is the radius of particle and $\epsilon_1$ is the dielectric
constant of the particles. Thus $\boldsymbol {\epsilon}^{(0)} =
\epsilon^{(0)}\boldsymbol{I}$ can be obtained from the (static)
Maxwell-Garnett formula:
\begin{equation}
{\epsilon^{(0)} - 1 \over \epsilon^{(0)} + 2} = \rho \left({\epsilon_1 - 1
\over \epsilon_1 + 2}\right), \label{MGA}
\end{equation}
where $\rho=4\pi a^3/3v$ is the volume fraction. Since {\bf p} can be
expressed in terms of ${\bf p}^{(0)}$, $\boldsymbol{\alpha}$ can be expressed
in terms of $\boldsymbol{\alpha}^{(0)}$
\begin{equation}
\label{eq:polariza} \boldsymbol{\alpha} =
{\boldsymbol{I}+\tau_0^2{\boldsymbol \omega_0}{\boldsymbol
\omega_0}+\tau_0({\boldsymbol \omega}_0 \times \boldsymbol{I})\over 1 +
(\omega_0 \tau_0)^2}\cdot \boldsymbol{\alpha}^{(0)}=\boldsymbol{A} \cdot
\boldsymbol{\alpha}^{(0)},
\end{equation}
where ${\boldsymbol \omega_0}{\boldsymbol \omega_0}$ represents a dyadic
tensor. It is worthy noting that the above derivation and the resulting
Eq.~\eqref{eq:polariza} are independent of coordinates and rotationally
invariant in space. Specifically, the most general form of $\boldsymbol{A}$
reads, for example, in the coordinate represented by the inset of
Fig.~\ref{fig:fig1}(b)
\begin{widetext}
\begin{equation}
\label{eq:A} {\boldsymbol A}={1\over 1+(\omega_0\tau_0)^2}\left[\left(
\begin{array} {c c c}
1+\omega_x^2\tau_0^2 & \omega_x\omega_y\tau_0^2 & \omega_x\omega_z\tau_0^2
\\
\omega_x\omega_y\tau_0^2 & 1+\omega_y^2\tau_0^2 & \omega_y\omega_z\tau_0^2
\\
\omega_x\omega_z\tau_0^2 & \omega_y\omega_z\tau_0^2 & 1+\omega_z^2\tau_0^2
\end{array}
\right)+ \left(
\begin{array} {c c c}
0 & -\omega_z\tau_0 & \omega_y\tau_0
\\
\omega_z\tau_0 & 0 & -\omega_x\tau_0
\\
-\omega_y\tau_0 & \omega_x\tau_0 & 0
\end{array}
\right)\right],
\end{equation}
\end{widetext}
where $\omega_0^2=\omega_x^2+\omega_y^2+\omega_z^2$. Thus, the
self-consistent equation [Eq.~\eqref{eq:DMGA}] becomes simultaneous equations
for the components of $\boldsymbol{\epsilon}$, which are generally quite
complicated.
\begin{figure}[b]
  \includegraphics*[clip, width=0.4\textwidth]{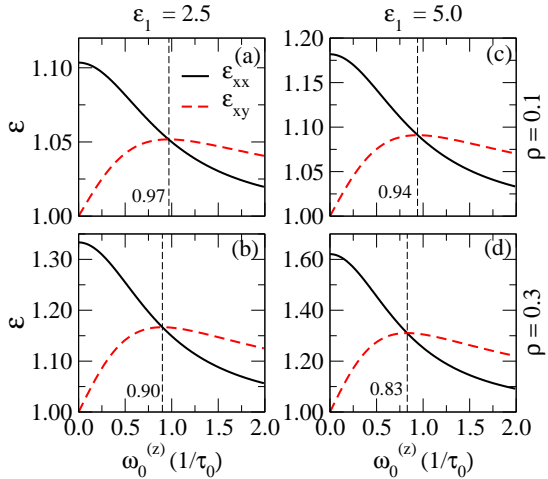}
  \caption{(Color online) The diagonal ($\epsilon_{xx}$) and
off-diagonal ($\epsilon_{xy}$) components of the effective
dielectric constant $\boldsymbol{\epsilon}$ for the case of uniform
rotation of angular velocity $\boldsymbol {\omega}_0=\omega_0 \hat
z$. Note that there is a maximum off-diagonal response where
$\epsilon_{xx}=\epsilon_{xy}$, and the peak position $\omega^*_0$
(marked by the vertical dashed lines) of $\epsilon_{xy}$ shifts to
the lower part as the volume fraction $\rho$ or the dielectric
contrast $\epsilon_1$ increases.}
  \label{fig:fig2}
  \end{figure}


\section{Special configuration in 2D}
\label{sec:2D} The solenoidal term consisting of the cross product
$\boldsymbol{\omega}_0 \times {\bf p}^{(0)}$ in the numerator of
Eq.~\eqref{eq:singlesolution} indicates a derogation of the dynamic effect in
the direction parallel to ${\boldsymbol \omega}_0$. Thus Eq.~\eqref{eq:DMGA}
can be discussed in a 2D case if ${\bf E}_0$ is parallel to one of
$\omega_x$, $\omega_y$, and $\omega_z$. In other words, one can always
decompose $\boldsymbol{\omega}_0$ to two parts, parallel and perpendicular to
${\bf E}_0$.  Due to this fact, it is instructive to simply consider a
special geometry in which $\boldsymbol {\omega}_0 = \omega_0 \hat{z}$, while
{\bf p} lies in the $xy$ plane, then Eq.~\eqref{eq:polariza} represents a
transformation $\boldsymbol{\alpha} = \boldsymbol{A}(\omega_0\tau_0) \cdot
\boldsymbol{\alpha}^{(0)}$, and
\begin{equation}
\label{eq:2D_A} \boldsymbol{A}(\omega_0\tau_0)=\left(
\begin{array} {c c c}
{1\over 1+(\omega_0\tau_0)^2} & -{\omega_0\tau_0\over 1+(\omega_0\tau_0)^2} &
0
\\
{\omega_0\tau_0\over 1+(\omega_0\tau_0)^2} & {1\over 1+(\omega_0\tau_0)^2} &
0
\\
0 & 0 & 1
\end{array}
\right),
\end{equation}
which indicates that the rotation does not affect the polarization along $z$
axis. In this case, the self-consistent equations become
\begin{subequations}
\label{eq:2DMGA}
\begin{eqnarray}
\epsilon_{xx}-1 &=& {\rho\over a^3}{\alpha^{(0)} \over
1+(\omega_0\tau_0)^2}(\epsilon_{xx}+2-\omega_0\tau_0
\epsilon_{xy}), \label{eq:diagonal}\\
\epsilon_{xy} &=& {\rho\over a^3}{\alpha^{(0)} \over
1+(\omega_0\tau_0)^2}[\epsilon_{xy}+\omega_0\tau_0
(\epsilon_{yy}+2)], \label{eq:offdiaognal} \\
\epsilon_{zz}-1 &=& {\rho\over a^3}\alpha^{(0)} (\epsilon_{zz}+2),
\label{eq:zz}
\end{eqnarray}
\end{subequations} while $\epsilon_{yy}=\epsilon_{xx},
\epsilon_{yx}=-\epsilon_{xy},
\epsilon_{yz}=\epsilon_{zy}=\epsilon_{xz}=\epsilon_{zx}=0$. Thus,
$\epsilon_{zz}$ is described by the usual MGA, as expected. Equations
\eqref{eq:2DMGA} suggest a complex notation. If we adopt complex notation
$\tilde{p}=p_x + i p_y$, the transformation
$\tilde{p}=\tilde{A}\tilde{p}^{(0)}$ can be achieved by a complex number
$\tilde{A}=1/(1-i\omega_0\tau_0)$. If we write
$\tilde{\epsilon}=\epsilon_{xx}+i\epsilon_{xy}$, the dynamic MGA
[Eq.~\eqref{eq:DMGA}] becomes a scalar but complex equation
\begin{equation}
{\tilde{\epsilon} - 1 \over \tilde{\epsilon} + 2} = \rho \tilde{A}
\left({\epsilon_1 - 1 \over \epsilon_1 + 2}\right). \label{eq:complex}
\end{equation}
One can show that Eqs.~\eqref{eq:DMGA} and \eqref{eq:complex} lead to exactly
the same self-consistent equations [Eqs.~\eqref{eq:diagonal} and
\eqref{eq:offdiaognal}] for this special configuration in 2D. The
polarization can be obtained from Eq.~\eqref{eq:P}
\begin{equation}
\label{eq:complexP} \tilde{P}={3\beta \over
4\pi(1-\beta-i\omega_0\tau_0)}={3\beta/(1-\beta) \over
4\pi(1-i\omega_0\tau)},
\end{equation}
where
\begin{equation}
\beta=\rho{\epsilon_1-1\over \epsilon_1+2}, \label{eq:beta}
\end{equation}
and
\begin{equation}
\tau={\tau_0 \over 1-\beta} \label{eq:tau}.
\end{equation}
As a consequence, when we vary the rotational angular velocity $\omega_0$,
$P_y \equiv \text{Im[}\tilde{P}\text{]}$ will exhibit a peak when
$\omega_0\tau=1$. These are clearly shown by Fig.~\ref{fig:fig2}, where we
plot the diagonal effective dielectric constant $\epsilon_{xx}$ (solid lines)
and the off-diagonal effective dielectric constant $\epsilon_{xy}$ (dashed
lines) as functions of $\omega_0$. The effective dielectric constant can be
solved directly from Eq.~\eqref{eq:DMGA}, Eq.~\eqref{eq:2DMGA}, or from
Eqs.~\eqref{eq:P} and \eqref{eq:complexP}. It is very interesting to see that
an electric field ${\bf E}_0=E_0{\hat x}$ can induce a polarization response
in the perpendicular direction $\hat {y}$, i.e., nonvanishing $P_y$ appears
due to the rotational motion $\boldsymbol {\omega}_0 = \omega_0 \hat{z}$ of
the particles, which accumulates part of the surface charges, or rotates the
dipole moment ${\bf p}$,  off the $\hat x$ direction. Due to this transfer by
rotational motion, the diagonal response $\epsilon_{xx}$ decreases
monotonically as the strength of the transfer effect increases (e.g.,
rotation speeds up), whereas a peak in $\epsilon_{xy}$ shows up. This
observation of off-diagonal response is quite interesting, resembling a
classic Hall effect. In Fig.~\ref{fig:fig2}, we also illustrate both the
effects of the volume fraction $\rho$ and the dielectric contrast
$\epsilon_1$ on the polarization responses, as well as on the peak shifting
of $\epsilon_{xy}$. We have specifically used $\epsilon_1=2.5$ in
Figs.~\ref{fig:fig2}(a) and \ref{fig:fig2}(b) (panels at the left), while
$\epsilon_1=5.0$ in Figs.~\ref{fig:fig2}(c) and \ref{fig:fig2}(d) (panels at
the right). Also we set $\rho=0.1$ and $0.3$ for the upper panels and the
lower panels, respectively. Combination of the effects of the dielectric
contrast $\epsilon_1$ and the volume fraction $\rho$ is actually captured by
the definition of $\beta$ in Eq.~\eqref{eq:beta}. It is clearly seen in
Fig.~\ref{fig:fig2} that increasing $\beta=0.033$, $0.057$, $0.1$, and
$0.171$ leads to a shifting of the peak position (vertical dashed lines) of
$\epsilon_{xy}$ at $\omega^*_0=0.97/\tau_0$, $0.94/\tau_0$, $0.90/\tau_0$,
and $0.83/\tau_0$. These correspond to Figs.~\ref{fig:fig2}(a),
\ref{fig:fig2}(c), \ref{fig:fig2}(b), and \ref{fig:fig2}(d), respectively.
\begin{figure}[t]
  \includegraphics*[clip, width=0.4\textwidth]{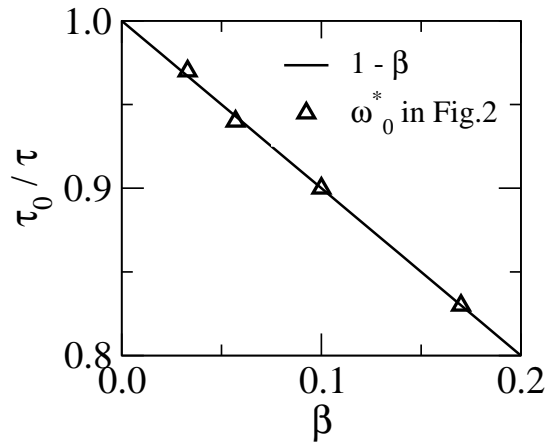}
  \caption{Interacting relaxation time ${\bf \tau}$
predicated by Eq.~\eqref{eq:tau} compares favorably to numerical
data extracted from Fig.~\ref{fig:fig2}.}
  \label{fig:fig3}
  \end{figure}
From another perspective, it is noticed from Eq.~\eqref{eq:tau} that
the relaxation time will increase from noninteracting relaxation tim
$\tau_0$ to $\tau$ due to the interactions. This is shown in
Fig.~\ref{fig:fig3}, in which the solid line represents
Eq.~\eqref{eq:tau} and the triangles ($\triangle$) are the data of
$\omega_0^*$ extracted from Fig.~\ref{fig:fig2}. Then we can regard
$\tau$ as the interacting relaxation time, and the maximum
off-diagonal response occurs when $\omega_0 \tau=1$, which is
equivalent to $\omega^*_0 \tau_0=1$. These results suggest a way to
measure the interacting relaxation time in the system.

To this end, we have demonstrated the off-diagonal response due to the
rotational motion. We would also like to look upon the geometrical meaning of
these results. It is easy to tell from Eq.~\eqref{eq:complexP} that
$\tan\phi=\omega_0 \tau$, where $\phi$ is the angle spanned by ${\bf E}_0$
and ${\bf E}_{\text{L}}$, as shown in Fig.~\ref{fig:fig1}(b). Also by some
simple calculations of the angles ($\phi$, $\theta$, and $\gamma$) shown in
Fig.~\ref{fig:fig1}(b) we find that
$\tan\theta=\tan(\phi-\gamma)=\tan\theta_{0}\equiv \omega_0\tau_0$, where
$\theta_{0}$ is the angle of induced dipole moment ${\bf p}$ with respect to
${\bf E}_0$ for the case of an isolated rotating particle at angular velocity
$\omega_0$. \cite{Yu05,WanPRe00} This is as expected and required by the
self-consistency of the theory, showing that this relationship still holds
for an individual rotating particle in the many-particle system: in
Fig.~\ref{fig:fig1}(b) the dipole moment ${\bf p}$ and the total field ${\bf
E}_{\text{total}}$ on this particle spans the angle $\theta=\theta_{0}$. Now
it is clear that the maximum off-diagonal response appears when $\phi=\pi/4$,
where $\omega_0=\omega^*_0$ indicating a resonance. Right at this point, the
parallel and perpendicular effective dielectric constant are equal, i.e.,
$\epsilon_{xx}=\epsilon_{xy}$, which is already demonstrated by
Fig.~\ref{fig:fig2}. In the limit of $\omega_0 \to 0$, $\epsilon_{xx}$ is
readily predicted by the static MGA [i.e., Eq.~\eqref{MGA}] and
$\epsilon_{xy}$ vanishes. This represents the case without rotation,
corresponding to $\phi \to 0$ and $\theta \to 0$ in Fig.~\ref{fig:fig1}(b).
Whereas in the opposite limit of $\omega_0 \to \infty$, both $\epsilon_{xx}$
and $\epsilon_{xy}$ approach unity (see also Fig.~\ref{fig:fig2}). This means
that the rotating particle lost its tensorial polarizability
$\boldsymbol{\alpha}$ of Eq.~\eqref{eq:polariza} [or seen from
Eq.~\eqref{eq:2D_A}] and the system dielectrically ``sees" no polarizable
suspension, i.e., the suspensions become ``invisible", because the very fast
rotational motion distributes the surface charges homogeneously along the
circumference of the particle, accumulating no net charges at any interfacial
place. In this case, both $\phi$ and $\theta$ lost definition because of
${\bf E}_{\text{L}}$=0. As a result, the system responses just as the pure
host media. It should be remarked that at a finite yet very fast rotation
$\omega_0 \gg 1/\tau_0$, $\phi \to \pi/2$ and $\theta \to 0$.
\begin{figure}[t]
  \includegraphics*[clip,width=0.4\textwidth]{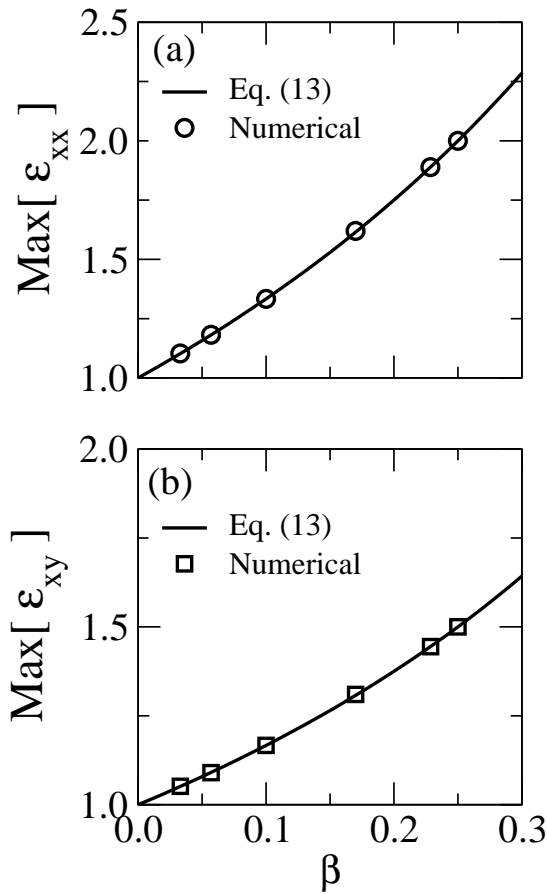}
  \caption{The dependence of the maximum polarization
responses on $\beta$. (a) $\epsilon_{xx}$ which occurs when
$\omega_0=0$, i.e., predicted by the static MGA, and (b)
$\epsilon_{xy}$ which occurs when $\omega_0=\omega^*_0$.}
  \label{fig:fig4}
  \end{figure}

Furthermore, we examine the maximum (e.g., denoted by Max[$\cdots$])
polarization responses of the system within the dynamic effective medium
theory, in terms of both diagonal $\epsilon_{xx}$ and off-diagonal
$\epsilon_{xy}$. It is quite straightforward from the previous discussion and
the data shown in Fig.~\ref{fig:fig2} that Max[$\epsilon_{xx}$] and
Max[$\epsilon_{xy}$] occur at $\omega_0=0$ and $\omega_0=\omega^*_0$
respectively, for a certain $\beta$. This is also explicitly indicated by
Eq.~\eqref{eq:complexP}. Open circles ($\bigcirc$) and squares ($\square$) in
Fig.~\ref{fig:fig4} show the Max[$\epsilon_{xx}$] and Max[$\epsilon_{xx}$]
extracted from a series figures like Fig.~\ref{fig:fig2} for various $\beta$,
while solid lines represent $\epsilon_{xx}(\omega_0=0)$ and
$\epsilon_{xy}(\omega=\omega_0^*)$. The good agreements further show the
usefulness of Eqs.~\eqref{eq:complex} and \eqref{eq:complexP} in the 2D case,
which appears simple but contains enough information in understanding the
interesting properties of the system. It is reasonable that both
Max[$\epsilon_{xx}$] and Max[$\epsilon_{xy}$] increase monotonically with
$\beta$, because commonly either an increased dielectric contrast or an
increased volume fraction lead to larger effective dielectric constant.

\section{Discussion and Conclusion}
To conclude, we have developed an dynamic MGA to take into account the
many-body interactions between rotating particles in a dynamic ER fluid. The
theory, like static MGA, stems from the application of the Lorentz local
field concept to a system of polarized particles in an external electric
field. The only difference is now that the particles are all rotating with
some angular velocity which leads to an off-diagonal polarization. We have
assumed uniform rotational angular velocity in the present paper. This
assumption may break down in shear flow under non-steady conditions. If the
angular velocity with the same magnitude $\omega_0$ distributes uniformly in
the solid angle, there is a reduction factor associated with the average
longitudinal (diagonal) response (parallel to ${\bf E_0}$) while the
transverse (off-diagonal) responses vanish identically. Namely, $\langle {\bf
p_{\parallel}} \rangle =  {\bf p}^{(0)}
[1-2\omega_0^2\tau_0^2/3(1+\omega_0^2\tau_0^2)]$ while $\langle {\bf
p}_{\perp} \rangle =0$, here $\langle \cdots \rangle$ represents average over
the solid angle.

Along with the present results, it is easy to derive a dynamic Bruggemann
effective medium approximation (EMA). In contrast to the present asymmetric
dynamic MGA, it is symmetrical and appropriate for two types of rotating
particles (e.g., of $\epsilon_1$ and $\epsilon_2$) whose volume fractions
satisfy $\rho_1 + \rho_2 = 1$. In the same sprit of Eq.~\eqref{eq:complex} in
the special 2D geometry, the self-consistent equation of dynamic EMA reads
\begin{equation}
\label{eq:EMA} \tilde{A}_1 \rho_1 {(\epsilon_1/\tilde{\epsilon}_e - 1) \over
(\epsilon_1/\tilde{\epsilon}_e + 2)} + \tilde{A}_2 \rho_2
{(\epsilon_2/\tilde{\epsilon}_e - 1) \over (\epsilon_2/\tilde{\epsilon}_e +
2)} = 0,
\end{equation}
where the two complex numbers $\tilde{A}_1$ and $\tilde{A}_2$ correspond to
the transformation coefficients of two types of rotating particles. These
complex numbers are assumed to be independent of the effective medium of
$\boldsymbol{\epsilon}_e$. Physically Eq.~\eqref{eq:EMA} means that the
induced dipole moment of type $\alpha$ (=1, 2) rotating particle in an
effective medium points to a different direction due to the relaxation of
surface charges. However, since the different types of particles are embedded
in an effective medium, the average dipole moment must vanish, thus yielding
the dynamic EMA which would be applicable to analogous problems like
magnetotransport in granular materials. These dynamic effective medium
theories [e.g., Eq.~\eqref{eq:DMGA} or Eq.~\eqref{eq:EMA}] provide guidelines
for numerical simulation on the dynamic effect in ER fluids \cite{Hu04} and
should be distinguished from those effective medium theories developed for
composites of anisotropic inclusions, \cite{aniso} bi-anisotropic media,
\cite{bian} and chiral media, \cite{chiral, chiral1} etc.

Extension to nonlinear ac responses \cite{Tian06_2,Tian06} is possible. In
which case, peak response can be found in parallel response of higher
harmonics. If we further apply an ac field at $\omega$ along the $x$ axis,
then there are two characteristic responses at the sum and difference of
frequencies: $\omega_0+\omega$, and $\omega_0-\omega$. When we vary $\omega$,
it can be shown that $P_y$ exhibits either peak at $\omega_0 + \omega_1$ or
$\omega_2 - \omega_0$. Thus, we can also determine the interacting relaxation
time $\tau$ by measuring the ac responses from the relations $(\omega_0 +
\omega_1)\tau=1$ or $(\omega_2 - \omega_0)\tau=1$. It is quite interesting
that the off-diagonal polarization shows a peak below $\omega_0$ in the
linear case while it peaks exactly at $\omega_0=1/\tau_0$ for isolated
rotating particle. So the ac field of $\omega$ can be applied perpendicular
or parallel to ${\bf E}_0$. The parallel component of ${\bf P}$ shows no peak
(it simply decrease as $\omega_0$ increases) in the linear case but its third
harmonic term as a function of $\omega_0$ has a peak. \cite{Tian06}

Finally, we would like to discuss several complications that may introduce
corrections to the local field and therefore the macroscopic responses (e.g.,
the effective dielectric constant $\boldsymbol \epsilon$). (1) since most ER
fluid might be anisotropic, e.g., somehow chains are formed in the direction
of applied electric field ${\bf E}_0$, the depolarization term may not be
isotropic. When the lattice symmetry is lowered by an external means---under
the influence of an external force/torque---the lattice is deformed, either
lengthened in one direction and/or contracted in the other direction.
\cite{LoJPC2001} These geometric anisotropy effect will introduce a
non-isotropic local field correction. \cite{HuangJPCB} This additional
anisotropy must also be accommodate in a consistent way. (2) The assumption
of isotropic depolarization is necessary because of rotational invariance.
More precisely, the Lorentz cavity is always spherical [see
Fig.~\ref{fig:fig1}(b)]. When ${\bf P}$ is rotated away from ${\bf E}_0$ due
to the competition between the displacement and relaxation of surface
charges, the Lorentz relation ${\bf E}_{\text L}=4\pi {\bf P}/3$ remains
unchanged. However, the detailed dynamical correlation between the rotating
particles may be further explored in presence of a distribution of rotational
speed or intrinsic relaxation time. (3) There may exist certain uniaxial
gradient in one direction in the system, \cite{Xiao06} for instance, more
particles aggregate in a specific part of the system. (4) We have not taken
into account the effect of Brownian motion. In the presence of an external
electric field, the Brownian motion would become weaker than the case of no
external electric field, which is caused to appear by the electric
polarization of the particle. However, competition between rotational
Brownian motion and particle fibrillation is also a concern if the particles
are small enough to ensure rotational diffusion.

\section*{Acknowledgments}

This work was supported by the Research Grants Council Earmarked Grant of the
Hong Kong SAR Government. J.P.H. acknowledges the financial support by the
Pujiang Talent Project (No. 06PJ14006) of the Shanghai Science and Technology
Committee, by the Shanghai Education Committee and the Shanghai Education
Development Foundation (``Shu Guang" project), by Chinese National Key Basic
Research Special Fund under Grant No. 2006CB921706, and by the National
Natural Science Foundation of China under Grant No. 10604014.


{}


\begin{thebibliography}{}
\bibitem{Block84}Block, H.; Kluk, E.; McConnell, J.; Scaife, B. K.
{\it J. Colloid Interface Sci.} {\bf 1984}, {\it 101}, 320.

\bibitem{Hemp91}Hemp, J. {\it Proc. R. Soc. London, Ser. A} {\bf 1991}, {\it 434},
297.

\bibitem{Lad88} Ladd, A. J. C. {\it J. Chem. Phys.} {\bf 1998}, {\it 88}, 5051.

\bibitem{Williams00} Wahed, A. K. El.; Sproston, J. L.; Williams, E. W.
{\it J. Phys. D: Appl. Phys.} {\bf 2000}, {\it 33}, 2995.

\bibitem{Hu04}Dassanayake, U. M.;  Offner, S. S. R.; Hu, Y.
{\it Phys. Rev. E}, {\bf 2004}, {\it 69}, 021507.

\bibitem{Zukoski91}Klingenberg, D. J.; Swol, F.; Zukoski, C. F.
{\it J. Chem. Phys.} {\bf 1991}, {\it 94}, 6160.

\bibitem{Ohsawa95}Negita, K.; Ohsawa, Y.
{\it Phys. Rev. E} {\bf 1995}, {\it 52}, 1934.

\bibitem{Negita04}Misono, Y.; Negita, K.
{\it Phys. Rev. E} {\bf  2004}, {\it 70}, 061412.

\bibitem{WanPRe00}Wan, J. T. K.; Yu, K. W.; Gu, G. Q.
{\it Phys. Rev. E} {\bf 2000}, {\it 62}, 6846.

\bibitem{WanPRE01}Wan, J. T. K.; Yu, K. W.; Gu, G. Q.
{\it Phys. Rev. E} {\bf 2001}, {\it 64}, 061501.

\bibitem{Yu05}Yu, K. W.; Gu, G. Q.; Huang, J. P.; Xiao, J. J.
{\it Int. J. Mod. Phys. B} {\bf 2005}, {\it 19}, 1163.

\bibitem{HuangJPCB}Cao, J. G.; Huang, J. P.; Zhou, L. W.
{\it J. Phys. Chem. B} {\bf 2006}, {\it 110}, 11635.

\bibitem{Negita04Rotation}Misono, Y.; Negita, K.
{\it Phys. Rev. E} {\bf 2004}, {\it 70}, 061412.

\bibitem{EOR}Gimsa, J.; Muller, T.; Schnellea, T.; Fuhr, G.
{\it Biophys. J.} {\bf 1996}, {\it 71}, 495.

\bibitem{EOR1}Becker, F. F.; Wang, X. B.; Huang, Y.; Pethig, R.; Vykoukal,
J.; Gascoyne, R. R. C. {\it Proc. Natl. Acad. Sci. U.S.A.} {\bf 1995}, {\it
92}, 860.


\bibitem{NegativeER}C\={e}bers, A.; Lemaire, E.; Lobry, L.
{\it Phys. Rev. E} {\bf  2000}, {\it 63}, 016301.

\bibitem{NEgative2}C\={e}bers, A. {\it Phys. Rev. Lett.} {\bf 2004}, {\it 92}, 034501.

\bibitem{Tao05}Tao, R.; Lan, Y. C. {\it Phys. Rev. E} {\bf 2005}, {\it 72}, 041508.

\bibitem{Quincke}Quincke, G. {\it Ann. Phys. Chem.} {\bf 1896}, {\it 59}, 477.

\bibitem{Jackson1975}Jackson, J. D. {\it Classical Electrodynamics};
Wiley: New York, 1975.

\bibitem{aniso}Levy, O.; Stroud, D. {\it Phys. Rev. B} {\bf 1997}, {\it 56}
8035.

\bibitem{bian}Weiglhofer, W. S.; Lakhtakia, A.; Michel, B.
{\it Micro. Opt. Tech. Lett.} {\bf 1997}, {\it 15}, 253.

\bibitem{chiral}Sihvola, A. H.; Lindell, I. V.
{\it Elect. Lett.} {\bf 1990},  {\it 26}, 118.

\bibitem{chiral1}Lakhtakia, A.; Varadan, V. K.; Varadan, V. V.
{\it J. Mater. Res.} {\bf 1993}, {\it8}, 917.

\bibitem{Tian06}Tian, W. J.; Huang, J. P.; Yu, K. W.
{\it Phys. Rev. E} {\bf 2006}, {\it 73}, 031408.

\bibitem{Tian06_2}Tian, W. J.; Huang, J. P.; Yu, K. W.
{\it Chem. Phys. Lett.} {\bf 2006} 427, 101.

\bibitem{LoJPC2001}Lo, C. K.; Wan, J. T. K.;  Yu, K. W.
{\it J. Phys.: Condens. Matter} {\bf 2001}, {\it 13}, 1315.

\bibitem{Xiao06}Xiao, J. J.; Yu, K. W.
{\it Appl. Phys. Lett.} {\bf 2006}, {\it 88}, 071911.

\end{thebibliography}
\end{document}